\documentstyle[12pt,a4,epsfig]{article}

\def\mtev#1{{$#1\,$TeV}}
\def\mgev#1{{$#1\,$GeV}}

\def\g5{\gamma_5}

\def\imcom{\hbox{Im}}

\def\journal#1&#2(#3)#4{{\unskip,~\sl #1\unskip~\bf\ignorespaces #2\unskip~\rm (19#3) #4}}
\def\jour#1&#2(#3)#4{{\unskip ~\sl #1\unskip~\bf\ignorespaces #2\unskip~\rm (19#3) #4}}\def\p{{\bf p}}

\begin{document}
\sloppy
\thispagestyle{empty}

\mbox{}
\begin{center}
{\LARGE\bf Theory Aspects of Double-Spin Asymmetries} \\

\vspace{2mm}
{\LARGE\bf in Proton-Nucleon Collisions}\\

\vspace{2em}
\large
Glenn A. Ladinsky
\\
\vspace{2em}
{\it  Michigan State University}
 \\
{\it Department of Physics \& Astronomy, East Lansing, MI  48824-1116
 U.S.A.}\\
\end{center}
%
\begin{abstract}
\noindent
The study of two spin asymmetries in hadron-hadron collisions
probes the details of fundamental particle interactions in ways
infeasible to machines with unpolarized collisions.
Within reach is how the proton spin is distributed among its
constituents through $\Delta G$ and $\Delta\bar{q}$.
Measuring couplings, furthering our understanding jet structure and
uncovering new physics are all among the possibilities
available at polarized colliders.
Spin physics may also deepen our understanding of higher
twist behavior and the transition between perturbative and
nonperturbative physics.
An overview of the double spin physics is presented in this report.
\end{abstract}
\footnote{Talk presented during the Workshop on the {\it Prospects of
Spin Physics at HERA} held at DESY-Zeuthen, 
Germany, 28-31 August 1995.}

\vspace{1.0cm}

\section{Introduction}
\label{sect1}
The study of two spin asymmetries in hadron-hadron collisions
probes the details of fundamental particle interactions in ways
infeasible to machines with unpolarized collisions.
Within reach is how the proton spin is distributed among its
constituents through $\Delta G$ and $\Delta\bar{q}$.
Measuring couplings, furthering our understanding jet structure and
uncovering new physics are all among the possibilities
available at polarized colliders.
Spin physics may also deepen our understanding of higher
twist behavior and the transition between perturbative and
nonperturbative physics.

In this talk I present an overview on theoretical aspects of double spin
asymmetries in proton-nucleon collisions.
This is a large topic, and I do not intend to cover
all the double spin physics involved with all conceivable machines.
For this reason, examples have been selected to demonstrate
various ideas, and
since the Relativistic Heavy Ion Collider (RHIC)
has been approved, most of the examples available
tend to be  from studies at RHIC energies ($\sqrt{s}=50-500\,$GeV).
For further details, the reader is directed to the proceedings
of the many conferences that have been held on spin
physics ({\it e.g.}, \cite{confone,conftwo})
or to the various review articles that appear in
the literature\cite{reviews}.

By manipulating the spins
of the two initial particles in the collision,
we exert a maximum in control
over the degrees of freedom within our reach.
Nevertheless, we are not required
to restrict ourselves to spins solely in the initial state.
Double spin asymmetries can also be obtained with one spin
in the initial state and one spin in the final state or
even with both spins taken from the final state.

Starting with some general comments, we look at spin physics
from a density matrix approach and see what asymmetries
probe parity violation and what asymmetries probe higher twist
effects.  Proceeding from the perspective of a factorized
cross section, we find that the spin contributions
can be isolated into three separate sources: the parton distributions,
the hard scatter and the final state fragmentation or decay. We proceed
to delve into the details one at a time.


\section{General Comments}
\label{sect2}

\subsection{Some Double Spin Asymmetries{\rm [4,39]}}
\label{sect21}

The first two asymmetries we examine are single spin asymmetries.
With momentum and longitudinal spin information
on only one of the particles in the process, we have
\begin{eqnarray}
A_L={d\sigma(+)-d\sigma(-)\over d\sigma(+)+d\sigma(-)},
\end{eqnarray}
where $d\sigma(+)$ ($d\sigma(-)$) represents the differential cross
section for right-handed (left-handed) helicity.
In terms of the corresponding helicity amplitudes, $M(+)$ and $M(-)$,
this asymmetry has the dependence,
\begin{eqnarray}
A_L\propto [|M(+)|^2-|M(-)|^2].
\end{eqnarray}
Since parity conservation demands $|M(+)|^2=|M(-)|^2$, a nonzero
value for $A_L$ signals parity violation.

With momentum and transverse spin information
on only one of the particles in the process, we have
\begin{eqnarray}
A_T={d\sigma^\uparrow-d\sigma^\downarrow\over
                     d\sigma^\uparrow+d\sigma^\downarrow},
\end{eqnarray}
where $d\sigma^\uparrow$ ($d\sigma^\downarrow$)
represents the differential cross section when the particle's
spin vector is directed in the upward (downward) transverse direction.
As can be seen from density matrix calculations,
this asymmetry is proportional to
the well known ``helicity flip'' amplitude:
\begin{eqnarray}
A_T\propto \imcom [M(+)M(-)^\dagger].
\end{eqnarray}
Perturbative QED and QCD yield tree level amplitudes that are real;
consequently, $A_T=0$ is expected at this level.  Imaginary
amplitudes can appear, however, from loop diagrams and higher twist
effects, resulting in nonzero $A_T$.

Moving on to the double spin asymmetries, we discuss four possibilities:
\begin{eqnarray}
A_{LL}={d\sigma(++)-d\sigma(+-)\over d\sigma(++)+d\sigma(+-)},
\qquad
A_{LL}^{PV}={d\sigma(++)-d\sigma(+-)\over d\sigma(++)+d\sigma(+-)},
\end{eqnarray}
\begin{eqnarray}
A_{TT}={d\sigma(\uparrow\uparrow)-d\sigma(\uparrow\downarrow)
           \over d\sigma(\uparrow\uparrow)+d\sigma(\uparrow\downarrow)},
\qquad
A_{TL}={d\sigma(\uparrow +)-d\sigma(\downarrow +)
                \over d\sigma(\uparrow +)+d\sigma(\downarrow +)},
\end{eqnarray}
where the first (second) index in $d\sigma$ indicates the polarization
of the first (second) particle whose spin we are monitoring.
In this case, the density matrix tells us the dependence on the
helicity amplitudes for the longitudinal asymmetries is
\begin{eqnarray}
A_{LL}\propto [|M(++)|^2-|M(+-)|^2]
\qquad\hbox{and}\qquad
A_{LL}^{PV}\propto [|M(++)|^2-|M(--)|^2].
\end{eqnarray}
As the label indicates, $A_{LL}^{PV}\ne 0$ results from parity violation.
For $A_{LL}$, there is no reason to expect $|M(++)|^2=|M(+-)|^2$ in
QED or QCD, even at tree level.  With this nonzero spin combination
understood from a perturbative standpoint, we have a useful tool for probing
less understood contributions to the cross section, like the parton
distribution functions.
The asymmetry $A_{TT}$, like $A_{LL}$, also gains nonzero values at
tree level in QED and QCD due to terms like Re$[M(++)M(--)^\dagger]$.
The asymmetry $A_{TL}$ on the other hand, is similar to $A_T$ in that
it requires nonzero imaginary portions to the amplitude or higher
twist effects to generate nonzero values.

Summarizing, we find that the asymmetries $A_{LL}$ and $A_{TT}$, with
their nonzero values at leading order (LO), can be used as a tool to
study the nonperturbative structure in the cross section. The
asymmetries $A_L$ and $A_{LL}^{PV}$ are sensitive to parity violation
effects, and the asymmetries $A_T$ and $A_{TL}$ are better suited for
probing higher twist.

\section{Factorization}
\label{sect3}

The factorization of the cross section into perturbative and nonperturbative
parts, which we frequently use in our studies of unpolarized collisions,
also follows through for expressing polarized cross sections\cite{jccfac}.
In the case of $A_{LL}$, where the two incoming hadrons are polarized,
we have\cite{reviews}
\begin{eqnarray}
A_{LL}=\sum_{ij}{1\over 1+\delta_{ij}}\int dx_A\,dx_B
[\Delta f_{i/A}(x_A,Q)\Delta f_{j/B}(x_B,Q) + (i\leftrightarrow j)]
\hat{a}^{ij}_{LL}/d\sigma
\label{eq:faca}
\end{eqnarray}
where $\Delta f$ is the parton helicity density and $\hat{a}^{ij}_{LL}$
is the parton level asymmetry.
We can also consider asymmetries resulting from spin effects in
fragmentation.  Here, we show the spin dependence in the cross
section as reflected by a density matrix factor\cite{chlfrg},
\begin{eqnarray}
E_C {d^3\sigma \over d^3p_C}=\sum_{abc} \int dx_A\, dx_B {dz\over z}
&&\rho_{\alpha\alpha^\prime} f_{a/A}(x_A,Q)f_{b/B}(x_B,Q) \nonumber \\
&&d\sigma_{\alpha\alpha^\prime\beta\beta^\prime}(a+b\rightarrow c+X)
\rho_{\beta\beta^\prime} D_{C/c}(z),
\end{eqnarray}
where $f$ is the unpolarized parton density, $D$ is the fragmentation
function and $\rho$ is the density matrix over helicities 
$\alpha,\alpha^\prime,\beta,\beta^\prime$.
Analogous formulas follow for the case of transverse spins\cite{jccfac,artru}.

From this perspective, we identify three sources for
spin effects:
\begin{description}
\item[\hspace{0.5in} $\hat{a}$]the hard scattering asymmetry
\item[\hspace{0.5in} $\Delta$D]the helicity dependent fragmentation functions
\item[\hspace{0.5in} $\Delta$PDFs]the parton helicity distributions
\end{description}
Next, we discuss spin in each of these parts.  For details the
reader is directed to the appropriate references.

\section{Getting the Spin Information}
\label{sect4}

\subsection{Accessing Asymmetries in the Hard Scattering}
\label{sect41}

In general, polarization can be used to measure the couplings or form factors
that govern the interactions in the hard scattering portion of the
cross section.

In Ref.~\cite{zprimes} we investigated what can be learned about the 
couplings for a new scalar or vector boson
produced in hadron-hadron colliders with or without polarizing the beams.
We studied this from the perspective of a lowest order
Drell-Yan calculation within the
resonance of the new particle, where interference effects are minimal.
We recapitulate some of that discussion here.

We saw that
one longitudinally polarized beam permits a study of parity violation
in the lowest order Drell-Yan process while
with two longitudinally polarized beams spin one bosons may be
distinguished from spin zero bosons through helicity conservation.
Scattering with longitudinal beams alone,
however, is insufficient for distinguishing between the scalar and
the pseudoscalar couplings or between the vector and the axial vector
couplings.
Given that we know the boson spin,
we also do not gain access to any new information on the
couplings of the scalar or vector boson when both initial beams are polarized
longitudinally as compared to the case when only one beam was
longitudinally polarized.

It is only with {\it two} transversely polarized beams that we were able to
distinguish between scalar and pseudoscalar couplings or between vector
and axial vector couplings.  Furthermore, we found that comparisons
where the transverse polarizations of the two colliding hadrons are
perpendicular to each other versus parallel or antiparallel provides
a probe of the $CP$ invariance of the couplings provided the effects are
large enough to be measured.
To have only one beam with some transverse polarization gains nothing.

So, Ref.~\cite{zprimes} showed, using Drell-Yan
production of new bosons as an example,
that to make a complete measurement of the couplings
for a new physics process requires the
use of both transverse and longitudinal polarizations for the hadron beams.
If we were to consider a more general process than Drell-Yan,
our conclusion would still hold---full polarization information is needed
to measure all couplings.

In general, studying the couplings can get
more involved than our example.  One may wish to examine other interactions
besides those considered here which may enter the effective theory, {\it e.g.},
through a chiral lagrangian approach.
Loop corrections can generate chromoelectric or chromomagnetic
moments, and if these loops are generated through weak boson
exchange, parity violation may be introduced\cite{kly,xly}.

Specific sources for nonzero values of $\hat{a}$ are often predicted in models
that introduce physics beyond the standard model.
One example appears in
technicolor models\cite{taxtec}, where a four-fermion contact
term in the interaction lagrangian displays a manifest parity
violation in its $\gamma_5$ dependence,
\begin{eqnarray}
{\cal L}_{4q}\propto \bar{\Psi}\gamma_\mu (1-\eta\gamma_5)\Psi
\bar{\Psi}\gamma_\mu (1-\eta\gamma_5)\Psi,
\end{eqnarray}
where $\Psi$ is a quark doublet and $\eta=0,\pm 1$.
At energies of \mgev{\sqrt{s}=500} Taxil and Virey compute asymmetries
ranging over a -5\% to +10\% range at high transverse momentum.

Supersymmetry is also replete with parity violating interactions\cite{susy}.
For example, the pseudoscalar couplings to quarks or neutralinos and  
the charged higgs boson couplings to quarks carry a
different left-handed and right-handed behavior.
This manifest parity violation can easily create deviations
from the expected asymmetries for ordinary particle production\cite{reviews}.

Taken in the light of what has been reviewed
in this discussion, polarization is an essential tool for studying the
interactions of new bosons in hadron-hadron colliders.

\subsection{Accessing Asymmetries in the final state}
\label{sect42}

As we discussed previously, many asymmetries tend to vanish when
we only keep track of one spin in the reaction.
With only one spin in the initial state, it falls to the final
state to provide us with the second spin to obtain a nonzero asymmetry.
To extract this spin information in the final state typically
requires interpreting the particle correlations that
result from particle decay or fragmentation.

\subsubsection{Self-Analyzing Decays}
\label{sect421}

Particle decay is useful for extracting a final state polarization
because many particles are ``self-analyzing,'' meaning that the
decay distributions reflect the polarization of the parent particle.
Among the particles that have been analyzed from this perspective are the 
$\tau$\cite{taus}, the $\Lambda$ and other hyperons\cite{hyperons}.

As an example, we can look at some of the theoretical work that has
been done regarding polarization in top quark decays.  It was shown in
Ref.~\cite{kly} that the polarization of the $t$ (and the $\bar t$)
can be self--analyzed from the decay 
$t \rightarrow b W^+\rightarrow bl^+ \nu_{l}$
($\bar t \rightarrow \bar b W^-\rightarrow \bar{b}l^- \bar \nu_{l}$).  In the 
the rest frame of the $t$ ($\bar{t}$) quark, 
the preferred moving direction of the $l^+$ ($l^-$) 
is along the direction of the boost; {\it e.g.},
the $l^+$ likes to follow the boost direction of the right-handed top quark 
and the $l^+$ likes to go opposite the boost direction of the left-handed 
top quark.
Because of the correlations, it is possible to
distinguish different polarization states of the $t \bar t$ pairs by
the energy distribution of the leptons $l^+$ and
$l^-$ or from their angular distributions\cite{peskin,chang,skane}.

\epsfysize=3.0in
\centerline{\epsfbox{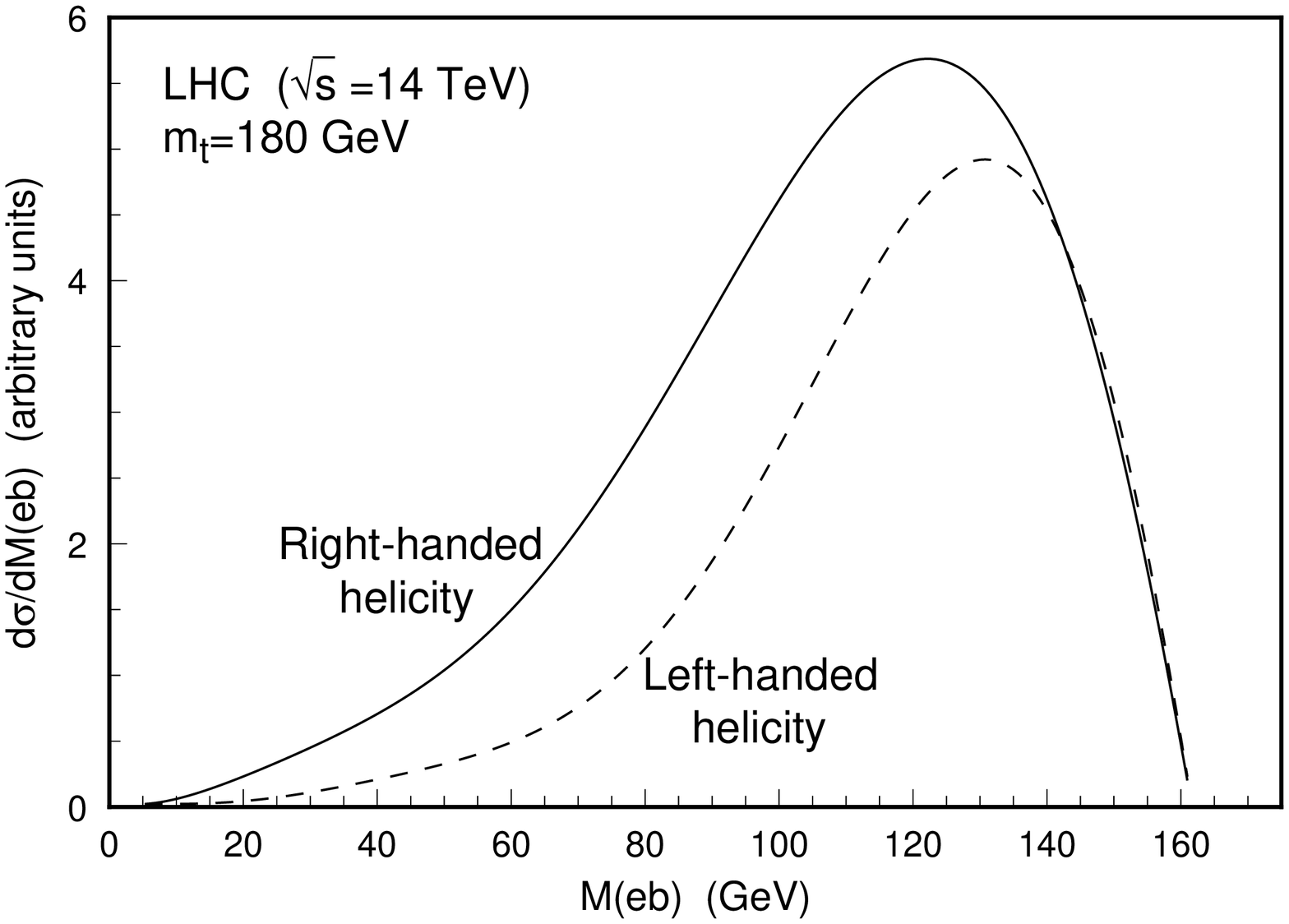}}

FIG.~1 These two curves represent the distribution $d\sigma/dM(eb)$ vs.
$M(eb)$ at the LHC for right--handed and left--handed top quark
helicities using $m_t=180\,$GeV and $m_b=0$.
Kinematic constraints in the lab frame on the rapidity ($|\eta|<2.5$)
and transverse momentum ($p_T>40\,$GeV) were imposed for the quarks and
visible leptons from the $t$ and $\bar{t}$ decays.

One important issue regarding the reality of analyzing polarization
through decay fragments appears when kinematic cuts are imposed 
on the data.  It is often the case that such constraints
bias the polarization of the sample.
To demonstrate the effect of the spin dependent decay,
we plot the $M(eb)$ distribution
for $q\bar{q} \to t\bar{t} \to b e^+ \nu \bar{b} q_1 \bar{q}_2$
at the LHC in Fig.~1, separating the contributions for left--handed
and right--handed helicities of the top quark.
($M(eb)$ is the mass of the $e^++b$ system.)
The difference between the two curves shows that it is necessary
in experiments and in Monte Carlo studies to keep track of the
effect kinematic constraints have with regard to the particle
polarizations\cite{argonne}.

\subsubsection{Jet Fragmentation}
\label{sect422}

Another analogous means for extracting spin information from
final state particles is to analyze the fragmentation products in 
jets\cite{efrall,jccbj}.
In the late 1970's, Nachtmann\cite{nacht} and Efremov\cite{efrfrg}
proposed looking at vector products of the momenta from selected
particles produced in the fragmentation.
In Ref.~\cite{nacht}, the vector product presented was
\begin{eqnarray}
T\equiv (\p_1\times\p_2)\cdot\p_3/|\p |,
\end{eqnarray}
where $\p$ is the jet momentum and the momenta for
the three leading particles in the jet obey $|\p_1|>|\p_2|>|\p_3|$.
One can then define the handedness of the jet by collecting the numbers
of events where $T$ is positive or negative,
\begin{eqnarray}
H={N(T<0)-N(T>0)\over N(T<0)+N(T>0)}=\alpha P,
\end{eqnarray}
where $P$ is the polarization of the jet and $\alpha$ is the analyzing power
we have for seeing the jet polarization.  The analyzing power must
be measured from experiments.
Measurements from the SLD experiment placed upper bounds on the size of
the analyzing power after observing results which they conclude
are consistent with zero.  Recent measurements by the DELPHI
collaboration have also found small analyzing powers\cite{efrdel}.

A correlated handedness, comparing the simultaneous fragmentation
between the two jets in $Z\rightarrow q\bar{q}$, has been found
by DELPHI of $8.5\%$.  This result is intriguing, not because
it is nonzero within errors, but because its
sign is opposite from standard model expectations\cite{efrdel}.

The fragmentation of transversely polarized quarks\cite{chlfrg,jccfrg,galjcc} 
will have an azimuthal distribution
\begin{eqnarray}
1+\alpha {{\bf s}_\perp\cdot({\bf t}\times\p )\over |\p_\perp|},
\end{eqnarray}
where the analyzing power here represents a left-right asymmetry,
${\bf s}_\perp$ is the transverse spin vector for the quark, and
$\p_\perp$ is the momentum of the observed particle
perpendicular to the jet axis, $\bf t$.
Providing that we understand the analyzing powers, we can use jet fragmentation
to pull spin information out of final state jets.

\subsection{Helicity Distributions and Structure Functions}
\label{sect43}

From Eq.~\ref{eq:faca}, we see that
our ability to extract the parton distribution functions
depends on how well we understand the hard scattering.
For this reason, processes which are understood through
perturbative calculations in the standard model are well
suited as tools that allow us to probe the helicity
dependence of the parton densities.

Helicity amplitudes and asymmetries for $2\rightarrow 2$ hard scatterings 
at tree level are prevalent in the literature\cite{reviews,gwu}.
Depending upon the subprocess, these parton level asymmetries
range from $-1$ to $+1$, {\it e.g.}, $q\bar{q}$ annihilation requires
quark and antiquark to have opposite helicities, resulting in
$\hat{a}_{LL}=-1$.  Transverse asymmetries appear as an
azimuthal variation in the amplitude.

The question to ask then, is whether we have sufficient sensitivity
to the asymmetries to extract the parton densities\cite{under}.
From DIS results on protons and neutrons, we have a good understanding
of the up and down quark densities, but the best measurements on
gluon and sea quark spin densities will come from hadron-hadron
collisions.

\subsubsection{Inclusive jet production}
\label{sect431}

In inclusive jet production, we have the advantage of
a high event rate which makes it easy to distinguish the
various possibilities for gluon and sea quark densities.
Depending upon the kinematics selected, we can focus on
initial states with either gluons or quarks.
In Chiappetta, {\it et al.}\cite{chia}, leading order $A_{LL}$ asymmetries
vary from zero to about 20\%, with a good separation between
large and negligible $\Delta G$ for $p_T<60\,$GeV in
pp collisions at \mgev{\sqrt{s}=500}.  The asymmetry drops for
\mgev{\sqrt{s}=16} to about 10-15\%, yielding a smaller distinction
in this case for \mtev{p_T<1}.  With new techniques for computing
one loop helicity amplitudes\cite{bern}, 
we can look forward to next to
order (NLO) results in the near future.

Recall, if parity is conserved, $A_L$ from an inclusive jet cross section
must vanish.  It is by getting spin information on two of the particles
(or momenta from more than one particle in the final state) that
the nonzero $A_{LL}$ survives for a parity conserving theory.
There has been a recent rise in the output of theoretical work 
investigating the potential of
extracting spin information from a final state jet.  If the analyzing powers
prove to be large enough, then it is reasonable to see how we might
gain information from double spin asymmetries involving one initial
state spin and one final state spin.  
This question was approached by Stratmann and Vogelsang\cite{aifstrat}, where
the asymmetries $A_{LL}^{if}$ and $A_{TT}^{if}$ were computed at
the parton level for inclusive jet production.  
At \mgev{\sqrt{s}=100}, they find significant variation in the
cross section due to $\Delta G$
and $\Delta s$ variations for \mgev{p_T<15}.

\subsubsection{Direct photon production}
\label{sect432}

Just as direct photon production has been useful in determining the
unpolarized gluon density\cite{tung}, 
direct photon production in polarized processes
should be useful for determining $\Delta G$.  By measuring both the
photon and jet from $qg\rightarrow q\gamma$, the RHIC should be able
to establish the $x$ dependence in $\Delta G(x)/G(x)$.
Higher order corrections have been computed for direct photon production
in polarized hadron collisions\cite{tkachov}.  
Contogouris, {\it et al.}, compute
$A_{LL}$ for \mgev{\sqrt{s}=38,100,500};  it is at the higher rapidities
and transverse momenta where they find the largest
variations in the asymmetry due to different
$\Delta G$ and $\Delta\bar{q}$, and it is also here that the
$K$-factors are largest.  The asymmetries get sizable,
up to 60\% at high transverse momentum.

\subsubsection{QCD-Electroweak Interference}
\label{sect433}

Jet production, even though dominated by QCD processes, may carry a parity
violating behavior due to the electroweak production mechanisms and
their interference with QCD processes\cite{taxtec,ranft}.
This parity violation can appear both in $A_L$ and $A_{LL}^{PV}$.
Without cuts, Refs.~\cite{taxtec,ranft} demonstrate
asymmetries around the percent level for $pp$ and $p\bar{p}$
collisions ranging from \mgev{\sqrt{s}=250-850}.
Variations in $A_{LL}^{PV}$ due to PDF uncertainties have been 
demonstrated in Ref.~\cite{taxtec} to get as large as $0.01$, but at large 
transverse momentum; with high luminosities, the RHIC 
should be able to distinguish between extreme variations caused
by varying the PDF.

\subsubsection{Heavy Quark Production}
\label{sect434}

As with unpolarized parton distribution functions\cite{berger}, it is expected
that open heavy flavor
production provides an optional probe of the gluon spin density.
Studies of heavy flavor production in polarized collisions have been
performed at LO\cite{hquark} 
and NLO\cite{karrob}.
At leading order, large values of $\hat a_{LL}$ may be obtained, near $-1$ at
large transverse momenta.
In Ref.~\cite{karrob}, however,
sizable cancellations were found between the LO and
NLO contributions in the spin dependent production of heavy quarks, 
leading to the conclusion that
a reliable extraction of polarized
parton densities in hadron collisions may be more dependent on radiative
corrections than has been observed with unpolarized parton densities.

\subsubsection{The Drell-Yan Process, Sea Quarks and Electroweak 
Boson Production}
\label{sect435}

The cross section for the production of $\mu^-\mu^+$ pairs of mass $M$
in the parton model is\cite{reviews}
\begin{equation}
{M^3{d^2\sigma}\over dMdx_F} ={8\pi\alpha^2\over 9}{x_1x_2\over x_1+x_2}
\sum_{a}e_a^2 [f_{a/A}(x_1)f_{\bar{a}/A}(x_2)+ (1\leftrightarrow 2)]
\end{equation}
making the Drell-Yan mechanism a prime probe of
the sea quark distributions.
The resultant asymmetry is directly proportional to the polarized
sea quark density,
\begin{equation}
A_{LL}\propto {
\sum_{a}e_a^2[\Delta f_{a/A}(x_1)\Delta f_{\bar{a}/B}(x_2)+ 
                                            (1\leftrightarrow 2)]
\over
\sum_{a}e_a^2 [f_{a/A}(x_1)f_{\bar{a}/B}(x_2)+ (1\leftrightarrow 2)]
}.
\end{equation}

An electroweak interference can also appear between the photon
and $Z$ boson.  This has been studied at RHIC energies by Leader
and Sridhar\cite{leader}.  Whether the variations in $A_{LL}$ due to changes
$\Delta G$ and $\Delta s$ will be useful depends upon the event rates at the
machine of interest.\cite{under}

Armed with an understanding of the unpolarized valence quark densities,
electroweak boson production will provide an important test of
the flavor breaking of the sea quark distributions\cite{sofw,chiaw}.
With their parity violating interations, the asymmetries in
the production of $W^\pm$ and $Z$ bosons can generate significant
values of $A_{LL}^{PV}$.
Bourrely and Soffer\cite{sofw} 
show that in $pp\rightarrow W^\pm +X$ for RHIC energies 
we can achieve asymmetries around 50\% with a large difference
appearing in the asymmetry for 
$pp\rightarrow W^-+X$ if $\Delta\bar{u}$ vanishes.

\subsubsection{Drell-Yan and Transverse Spin}
\label{sect436}

The transverse spin of partons in the proton is unmeasurable in the case of 
deep inelastic scattering\cite{dypire},
and due to helicity arguments, the transversity distribution for the
gluons in the proton is zero\cite{artru}.  
Nonetheless, with both incoming hadrons carrying transverse polarization, 
the Drell-Yan process is useful for studying the transversity distributions of 
valence and sea quarks\cite{dypire}.
Recall, by separating the azimuthal dependence of the cross into quadrants,
an asymmetry in the cross section is obtained,
\begin{equation}
{d\Delta\sigma\over dQ^2dyd\Omega}=
{4\alpha^2e^2\over 9 Q^2 s}h_T^A(x_A)\bar{h}_T^B(x_B) +(A\leftrightarrow B),
\end{equation}
where $h_T$ and $\bar{h}_T$ are distributions that measure the leading twist 
transverse polarization.
A study of the Drell-Yan cross section, 
with either photon or $Z$ boson intermediates, 
would complement an inclusive jet study.

\subsubsection{Drell-Yan and Higher Twist}
\label{sect437}

One longitudinal beam against one transverse beam accesses higher twist
in $pp\rightarrow \mu^-\mu^+ +X$.  Maintaining the intrinsic transverse
momentum dependence, Mulders and 
Tangermann\cite{mulders} find the asymmetry is
\begin{equation}
A_{LT}={\sin 2\theta\cos\phi\over 1+\cos^2\theta }
{\bar{U}^{LT}_{2,1}\over \bar{W}_T},
\end{equation}
where $\bar{W}_T = {1\over 3}\sum_{a}e_a^2 [f_{q/A}(x_1)f_{\bar{q}/A}(x_2)]$
and the higher twist contribution is contained in $\bar{U}^{LT}_{2,1}$.
Experimental results are needed to understand the size of this
asymmetry beyond a model dependence.

\subsubsection{Miscellaneous}
\label{sect438}

It should go without saying that there are a plethora of processes\cite{gwu}
that have nonzero double spin asymmetries, many of which have
been examined theoretically.
Among them are the production of 3-jets\cite{cjet}, 4-jets\cite{djet}, 
two photons\cite{bphoton}, 
J/Psi\cite{jpsi}, etc.
These processes have a variety of uses, but they usually take a back seat
to the processes with higher event rates and larger asymmetries.
These processes are useful for checking
the more precise methods of determining $\Delta G$ and $\Delta\bar{q}$.

We can also look forward to getting the one loop helicity amplitudes
to many processes through the techniques developed from sting 
theory\cite{bern}.

\section{Conclusions}
\label{sect5}

We have seen how the spin dependence of hadronic interactions can be
induced through the spin dependence of the parton distribution function,
the hard scatter, or the final state decay and fragmentation.
Double spin asymmetries in proton-nucleon collisions will fill and important
gap in our knowledge of how the proton spin is distributed among its
constituents by providing $\Delta G$ and $\Delta\bar{q}$.
These asymmetries also can provide an excellent means for understanding
fundamental interactions and facilitate studies of new physics.
Spin physics will pave the way to further our understanding of higher
twist and the transition between the perturbative and the
nonperturbative physics.  We may also have some surprises unfold
as we investigate variables that measure correlations in
jet fragmentation.

\section*{Acknowledgments}
\label{sect6}

My gratitude goes to Wolf-Dieter Nowak and Johannes Blumlein
for inviting me to DESY-Zeuthen and for their hospitality 
during my stay.
This work is funded in part by DESY-Zeuthen,
Michigan State University and NSF grant PHY-9309902.


\newpage

\end{document}